\documentclass[twoside]{JHEP3}
\title{KK-Masses in Dipole Deformed Field Theories}
\author{Karl Landsteiner and Sergio Montero\\
Instituto de F\'{\i}sica Te\'orica, C-XVI Universidad Aut\'onoma de Madrid\\
28049 Madrid, Spain\\
E-mail: \email{Karl.Landsteiner@uam.es, Sergio.Montero@uam.es}}
\abstract{We reconsider aspects of non-commutative dipole deformations of field
theories. Among our findings there are hints to new phases with spontaneous
breaking of translation invariance (stripe phases), similar to what happens in
Moyal-deformed field theories. Furthermore, using zeta-function regularization,
we calculate quantum corrections to KK-state masses. The corrections coming from
non-planar diagrams show interesting but non-universal behaviour. Depending on
the type of interaction the corrections can make the KK-states very heavy but
also very light or even tachyonic. Finally we point out that the dipole
deformation of QED is not renormalizable!}
\preprint{IFT-UAM/CSIC-06-03\\ \texttt{hep-th/0602035}}
\usepackage[english]{babel}
\usepackage{amssymb}
\usepackage{amsfonts}
\usepackage{amsmath}
\usepackage{graphicx}
\usepackage{epsfig}
\usepackage{axodraw}
\usepackage{psfrag}
\usepackage{cite}
\def\erf#1{(\ref{#1})} 
\newcommand{\cA}{{\cal A}}

\newcommand{\cM}{{\cal M}}  \newcommand{\cN}{{\cal N}}
\newcommand{\cO}{{\cal O}}  
  \newcommand{\cR}{{\cal R}}

  \newcommand{\bbR}{{\mathbb R}}

  \newcommand{\bbZ}{{\mathbb Z}}

\newcommand{\bR}{{\mathbf R}}  \newcommand{\bS}{{\mathbf S}}

\def\bsk{\boldsymbol{k}} \def\bsp{\boldsymbol{p}}

\newcommand{\be}{\begin{equation}}     \newcommand{\ee}{\end{equation}}
\newcommand{\bea}{\begin{eqnarray}}    \newcommand{\eea}{\end{eqnarray}}
\newcommand{\beann}{\begin{eqnarray*}} \newcommand{\eeann}{\end{eqnarray*}}
\newcommand{\bfig}{\begin{figure}}     \newcommand{\efig}{\end{figure}}
\newcommand{\ba}{\begin{array}}        \newcommand{\ea}{\end{array}}
\newcommand{\bcen}{\begin{center}}     \newcommand{\ecen}{\end{center}}
\newcommand{\btab}{\begin{tabular}}    \newcommand{\etab}{\end{tabular}}
\newcommand{\nn}{\nonumber}
\def\U{\operatorname{U}}

\def\tr{\operatorname{tr\:}}

\renewcommand{\Re}{\mathop{\rm Re}}   
\newcommand{\dd}{{\rm d}}
\newcommand{\e}{{\rm e}}
\newtheorem{Proposition}{Proposition}[section]

\newtheorem{Theorem}{Theorem}[section]
\newtheorem{Lemma}{Lemma}[section]
\newtheorem{Corrolary}{Corrolary}[section]

\newcommand{\bp}{\begin{Proposition}}	\newcommand{\ep}{\end{Proposition}}
\newcommand{\bt}{\begin{Theorem}}	\newcommand{\et}{\end{Theorem}}
\newcommand{\bl}{\begin{Lemma}}		\newcommand{\el}{\end{Lemma}}
\newcommand{\bc}{\begin{Corrolary}}	\newcommand{\ec}{\end{Corrolary}}
\begin{document}

\section{Introduction}  \label{sec:intro}
In recent years, a huge amount of work has been devoted to the study of
non-commutative field theories.
The main focus has been the Moyal bracket case, based on a non-trivial
commutator of coordinates
$[x^\mu, x^\nu ] = i \Theta^{\mu\nu}$. Part of the interest in these theories is
triggered by the
observation that they are realized in string theory on the world volume of
D-branes in a B-field background. The B-field is an antisymmetric second rank
tensor, and the Moyal bracket deformation is obtained by choosing a polarization
of the B-field with both indices along the directions of the world volume of the
D-brane. It is also possible to arrange the B-field in a different way, with one
index along the brane directions and the other one transverse to them. Also in
this case one obtains a deformation of the world volume theory. The deformation
in question is defined by the star-product
\bea
\phi_1(x) \star \phi_2(x) &:=& \exp\Big[ -\frac{1}{2} (L_2^\mu \partial^x_\mu -
L_1^\mu \partial^y_\mu)\Big] \,\phi_1(x) \,\phi_2(y)\Big|_{x=y} \nonumber \\
&\,=& \phi_1\left( x-\frac{L_2}{2} \right) \phi_2 \left( x + \frac{L_1}{2}
\right) ~, \label{eq:star.product}
\eea
which was first constructed in \cite{Bergman:2000cw} by considering T-duality of
Moyal-bracket deformed theories, and its basic field theoretical properties were first studied in
\cite{Dasgupta:2001zu}. As explained there, $L^\mu_{1,2}$ are the so-called dipole lengths of the
fields $\phi_1$ and
$\phi_2$. 
Associativity of this star-product
\be
(\phi_i \star \phi_j) \star \phi_k = \phi_i \star (\phi_j \star \phi_k) ~,
\ee
demands that the dipole length of a product of fields be the sum of the dipole
length of each field
\be
L_{\phi_1\star\cdots\star\phi_n} = L_{\phi_1} +\ldots +L_{\phi_n} ~,
\ee
i.e. the dipole length is \textit{additive}. Supergravity backgrounds dual to dipole
deformed field theories have been investigated in
\cite{Bergman:2001rw,Dasgupta:2000ry,Alishahiha:2002ex,Ganor:2002ju,
Alishahiha:2003ru} whereas various field theory aspects have been described in
\cite{Sadooghi:2002ph,Correa:2002cd,Chiou:2003sh}. It is
important to note that these dipole lengths are always related to $\U(1)$
symmetries of the (undeformed) field theory under consideration. Without
recourse to string theory, a dipole deformation of a field theory can be defined
by introducing the dipole lengths of the fields according to
\begin{equation}
L_\phi^\mu = \ell^\mu_a Q^a_\phi ~,
\end{equation}
where $Q^a_\phi$ are $\U(1)$ charges of the field $\phi$ and the matrix
$\ell^\mu_a$ picks out a certain linear combination. The Lagrangian of the
deformed field theory is then obtained by simply multiplying the fields with the
star-product (\ref{eq:star.product}). Since the terms in the Lagrangian are
neutral under the chosen $\U(1)$ symmetries, it is clear that the dipole lengths
of all the fields in a Lagrangian term add up to zero. In this case one is
allowed to delete one star from the product under the integral, i.e.
\begin{equation}
\int \dd x\; \phi_1(x) \star \phi_2(x) \star \dots \star \phi_n(x) = \int \dd x
\; \phi_1(x) \,\phi_2(x) \star \dots \star \phi_n(x) ~.
\end{equation}
An immediate consequence is that the quadratic terms in the Action remain
undeformed, and therefore
the propagators in the deformed quantum theory are the same as those in the
undeformed one. 
The integral defines the trace on the \mbox{$C^*$-algebra} of functions defined
by the deformed product. In the case of matrix valued fields this also includes
a trace on the matrix indices. However, the necessary cyclicity condition is
only fulfilled if the total dipole length of the integrand is \textit{zero},
because
only then one can delete one star under the integral and therefore permute the
fields cyclically.

Since the dipole deformation is based on the presence of $\U(1)$ charges, it is
clear that the
dipole moment of neutral fields is zero. For example, there is no non-trivial
dipole deformation of real scalar field theory and for the same reason the
dipole length of gauge fields has to vanish. In the case of complex fields,
demanding that
\be
(\phi^\dag \star \phi)^\dag =\phi^\dag \star \phi ~,
\ee
shows that $L_{\phi^\dag} =-L_{\phi}$. 

Although this structure is very similar to the well-known Moyal bracket
deformation of field theories, it has been relatively little investigated.
However, recently a very interesting proposal has been put forward by N\'u\~nez
and G\"ursoy. They considered the supergravity background as a specific dipole
deformation of the theory living on D5-branes wrapped on an $S^2$ inside a
Calabi--Yau, in such a way as to preserve $\cN=1$ supersymmetry in the four
dimensional flat part of the world volume. Such supergravity duals of confining
gauge theories are in general plagued by a rather unwelcome feature: the scale
of the masses of the KK-states coming from the compactified part of the world
volume is of the same order as the scale of the four dimensional gauge theory of
interest. Therefore, one cannot disentangle the interesting strongly coupled
gauge theory dynamics from the artefacts of these KK-states.

N\'u\~nez and G\"ursoy pointed out that this situation might be improved if one
considers a dipole deformed D5-brane theory \cite{Gursoy:2005cn}. More
specifically, using the techniques developed in \cite{Lunin:2005jy} they
constructed the supergravity background of such a theory in a B-field background
with one index of the B-field along the $S^2$ and another one transverse to the
D5-brane volume. They noted then that the volume of the compact internal
manifolds in the deformed background are smaller than in the undeformed one,
therefore indicating a possible disentanglement of the KK-states from the
interesting gauge theory dynamics. Further aspects of KK-states in these
supergravity backgrounds have been discussed in \cite{Bobev:2005ng, Pal:2005nr}.

This work motivated us to investigate the issue of KK-state masses in dipole
deformed theories from the
purely field theoretical point of view. Of course, the underlying theory
\cite{Gursoy:2005cn} is very complicated, and since its UV completion ultimately
is little string theory, it is not really a field theory. We will therefore
study much simpler examples of dipole deformations of  field theories
compactified on a circle. We found however that even the uncompactified theories
show a rather interesting behaviour, that so far has not been discussed in the
literature. Therefore, we will start in section two by investigating dispersion
relations in dipole deformed scalar field theories in three, four and five
dimensions. We will find indications for a phase transition at a certain
critical dipole length in three dimensions and five dimensions, whereas the four
dimensional theory is free of such behaviour at least in the weak coupling
regime.

In section two we move on to study the same scalar field theory compactified on
a circle. We find that the two-point amplitudes can be expressed in terms of
Epstein zeta functions. We will phrase the discussion in the language of thermal
field theory by introducing the inverse circumference $T=1/(2 \pi R)$.
Interestingly, the amplitudes depend on the dipole length only through the
non-integer part of the product $T L$ and, for $TL$ being close to an integer,
the non-planar two-point amplitudes grow without bound. Depending then on the
type of interactions the non-planar corrections to the KK-masses can be very
large and of positive or negative sign. In the latter case this indicates a
tachyonic instability (this is very similar to the behaviour of non-commutative
Moyal bracket deformed theories compactified on a non-commutative torus
\cite{Guralnik:2002ru}).

In section three we investigate the dipole deformation of the massless
Wess--Zumino model. The undeformed model is supersymmetric but, since the
$\U(1)$ symmetry that we use to define the dipole moments is the $\cR$-symmetry,
supersymmetry is broken in the deformed theory. We show that the leading
divergences appearing in the planar graphs still cancel whereas the
corresponding (leading) $L$ dependence of the non-planar amplitudes do not
cancel.

In section four we have a very brief look to dipole deformed QED. We use the
$\U(1)$ gauge symmetry for the dipole deformation. It turns out that already the
one-loop corrections to the polarization tensor
give rise to momentum dependent divergences that can not be absorbed in a
controlled way in the parameters of the tree level Lagrangian, therefore the
model is not renormalizable!

We conclude with a summarizing discussion in section five. In the appendix we
outline the analytic continuation of the Epstein zeta function and provide some
details of the calculations of section \ref{sec:wzmodel}.

\section{The bosonic dipole theory}  \label{sec:bosonic}
Following the preceding section, we begin by formulating a dipole theory for
complex scalar fields $\phi$ and $\phi^\dag$ with quartic interactions. The
action is
\be
S =\int\dd^D x ~\Big( \partial_\mu\phi^\dag \partial^\mu\phi -m^2\phi^\dag\phi
-\lambda(\phi^\dag\star\phi\star\phi^\dag\star\phi)
-g(\phi^\dag\star\phi^\dag\star\phi\star\phi) \Big)(x) ~.
\ee

As said, we have deleted the star-product in the quadratic terms. The bare
propagator is thus as in the commutative theory
\be
\widetilde\Gamma^{(2)}_{\phi\phi^\dag,\rm bare}(k) =\frac{i}{\bsk^2 -m^2
+i\varepsilon} ~.
\ee
There are two possible orderings of the fields in the interaction vertex. In the
first one with coupling $\lambda$, the star-product shifts the arguments of all
fields in the same direction and by the same amount. Therefore it can be undone
by a compensating shift of the integration variable. Thus, the
$\lambda$-coupling gives the same interactions as in the undeformed theory.
Only the second term with coupling $g$ produces a new form of the interaction.
In momentum space it is
\bea
S_{g} &=& -g \int\dd^D x ~(\phi^\dag\star\phi^\dag\star\phi\star\phi)(x)
=\ldots= \nn \\
&=& -g \,(2\pi)^D \int \left( \prod_{i=1}^4 \frac{\dd^D \bsk_i}{(2\pi)^D}
\right) \delta^D(\Sigma \bsk_i) \times \nn \\
&& \hspace*{5em} \times \e^{-iL_\phi \cdot(\bsk_1-\bsk_2-\bsk_3+\bsk_4)/2}
~\widetilde\phi^\dag(\bsk_1) \,\widetilde\phi^\dag(\bsk_2)
\,\widetilde\phi(\bsk_3) \,\widetilde\phi(\bsk_4) ~,
\eea
and so one obtains the vertex for the $g$-term in Minkowski space $\bR^{1,D-1}$
\vspace*{-3ex}
\be
\begin{picture}(50,50)(0,23)
\DashArrowLine(0,50)(25,25){3} \DashArrowLine(0,0)(25,25){3}
\DashArrowLine(50,50)(25,25){3} \DashArrowLine(50,0)(25,25){3}
\Text(4,37)[c]{$k_1$} \Text(3,14)[c]{$k_2$} \Text(47,13)[c]{$k_3$}
\Text(47,37)[c]{$k_4$}
\Text(-8,50)[c]{$\phi^\dag$} \Text(-8,0)[c]{$\phi^\dag$} \Text(57,0)[c]{$\phi$}
\Text(57,50)[c]{$\phi$}
\end{picture} \qquad = -ig \,\exp\left( -\frac{i}{2}L_\phi
\cdot(\bsk_1-\bsk_2-\bsk_3+\bsk_4) \right) ~.
\ee
\vspace*{2ex} \\
We also note that in the case $g = - \lambda$ the interaction is given by the
square of the
$\star$-product commutator $[ \phi^\dagger , \phi]$.

\subsection{The uncompactified theory}  \label{ssec:tachyon}
\begin{figure}[htbp]
\centering
\begin{picture}(200,75)(0,-5)
\DashArrowLine(0,50)(25,25){3} \DashArrowLine(0,0)(25,25){3}
\DashLine(50,50)(25,25){3} \DashLine(50,0)(25,25){3}
\DashCArc(25,50)(25,0,180){3}
\Text(4,37)[c]{$k$} \Text(4,14)[c]{$p$}
\Text(-8,50)[c]{$\phi^\dag$} \Text(-8,0)[c]{$\phi^\dag$} \Text(57,0)[c]{$\phi$}
\Text(57,50)[c]{$\phi$}
\DashArrowLine(150,50)(175,25){3} \DashArrowLine(150,0)(175,25){3}
\DashLine(200,50)(175,25){3} \DashLine(200,0)(175,25){3}
\DashCurve{(150,50)(187,40)(200,0)}{3}
\Text(154,37)[c]{$k$} \Text(154,14)[c]{$p$}
\Text(142,50)[c]{$\phi^\dag$} \Text(142,0)[c]{$\phi^\dag$}
\Text(207,0)[c]{$\phi$} \Text(207,50)[c]{$\phi$}
\end{picture}
\caption{The deformed vertex allows for a planar and a non-planar one-loop
correction to the two-point function.}
\label{fig:planar.vs.non-planar}
\end{figure}
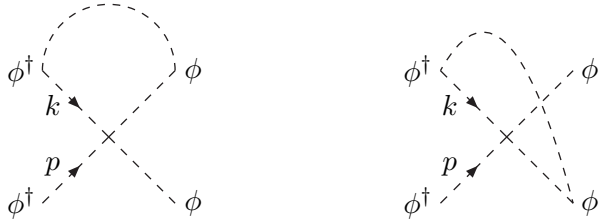

As shown in figure \ref{fig:planar.vs.non-planar}, the deformed vertex allows
for two inequivalent one-loop corrections to the two-point function. There is a
planar graph in which all the phases cancel, and a \mbox{non-planar} one in
which the phase depends on the loop momentum $\bsk$ as well as on the external
momentum $\bsp$. Since the planar amplitude is independent of the external
momentum we can absorb it in an infinite renormalization of the mass or set it
to zero by using dimensional regularization. The non-planar amplitude is given
by
\begin{equation}
-i \cA_{\rm n-p} = 2 g \cos(\bsp\cdot L) \int \frac{\dd^D \bsk}{(2\pi)^D}
\;\frac{\e^{i \bsk\cdot L}}{\bsk^2 - m^2+i\epsilon} ~.
\end{equation}
The amplitude can be easily evaluated by Wick rotating to Euclidean signature
and introducing a Schwinger parametrization
\begin{eqnarray}
\cA_{\rm n-p} &=& 2 g \cos(\bsp_E \cdot L_E) \int\dd \alpha \int \frac{\dd^D
\bsk_E}{(2\pi)^D} \;\exp \Big( -\alpha (k_E^2 + m^2) + i \bsk_E \cdot L_E \Big)
= \\
&=&  2 g \cos(\bsp_E \cdot L_E) \int\dd\alpha \left(\frac{1}{4 \pi \alpha}
\right)^{D/2} \exp \left( -\alpha m^2 -\frac{L^2}{4\alpha} \right) =\\
&=& 2 g \cos(\bsp_E \cdot L_E) \,\frac{m^{D-2}}{\sqrt{(2\pi)^D}}
\,\frac{K_{D/2-1}(|L m|)}{|L m|^{(D-2)/2}} ~,
\end{eqnarray}
where $K_n(x)$ is a modified Bessel function of the second kind. The UV
divergence is regulated in the non-planar graph by the dipole length. The
leading behaviour for $Lm \rightarrow 0$
is 
\begin{equation}
\cA_{\rm n-p} = \frac{g}{2 \pi^{D/2}} \,\cos(\bsp_E \cdot L_E) \,\Gamma\left(
\frac{D-2}{2} \right) L^{2-D} ~.
\end{equation}
This gives rise to a modified dispersion relation of the form
\begin{equation}
E^2 = p^2 + \frac{g}{2 \pi^{D/2}} \,\cos(p L) \,\Gamma\left( \frac{D-2}{2}
\right) L^{2-D} ~,
\label{eq:dispersion}
\end{equation}
where for simplicity we took the massless limit and also considered the $D-1$
momentum $\vec{p}$ to be parallel to $L$. This in particular implies that $L$ is
a spacelike vector and that there is a coordinate system in which $L$ has
non-vanishing component only in the $D$-th direction $L_\mu =(0,0,\dots,L)$.
 
\EPSFIGURE{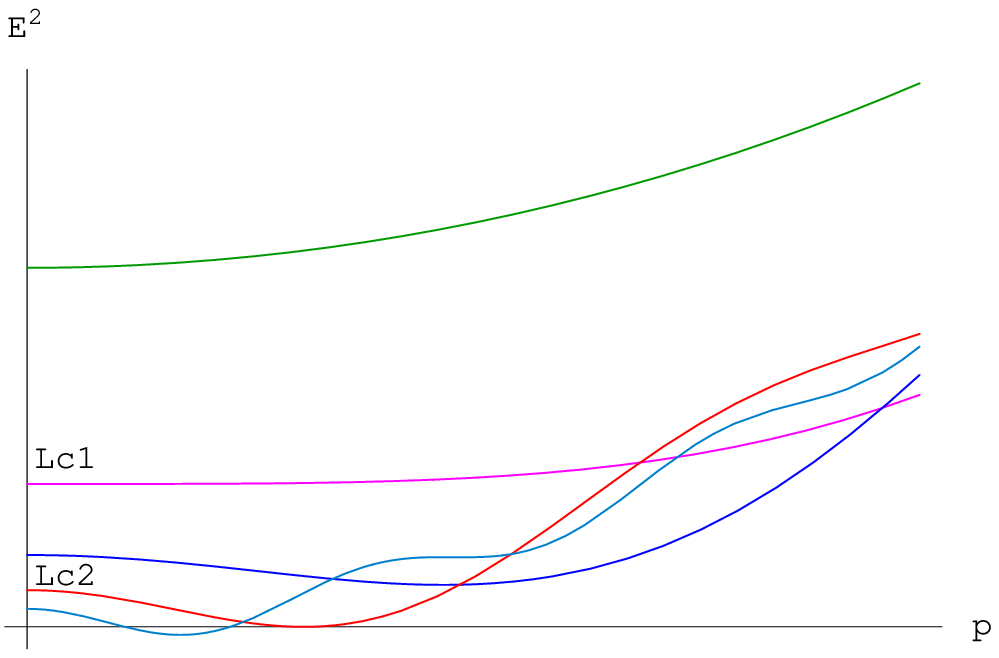}{Dispersion relation for different dipole lengths
in $D=3$. At $L_{c1}$ it develops a minimum away from $p=0$ and at $L_{c2}$ it
touches $E=0$.}
\EPSFIGURE{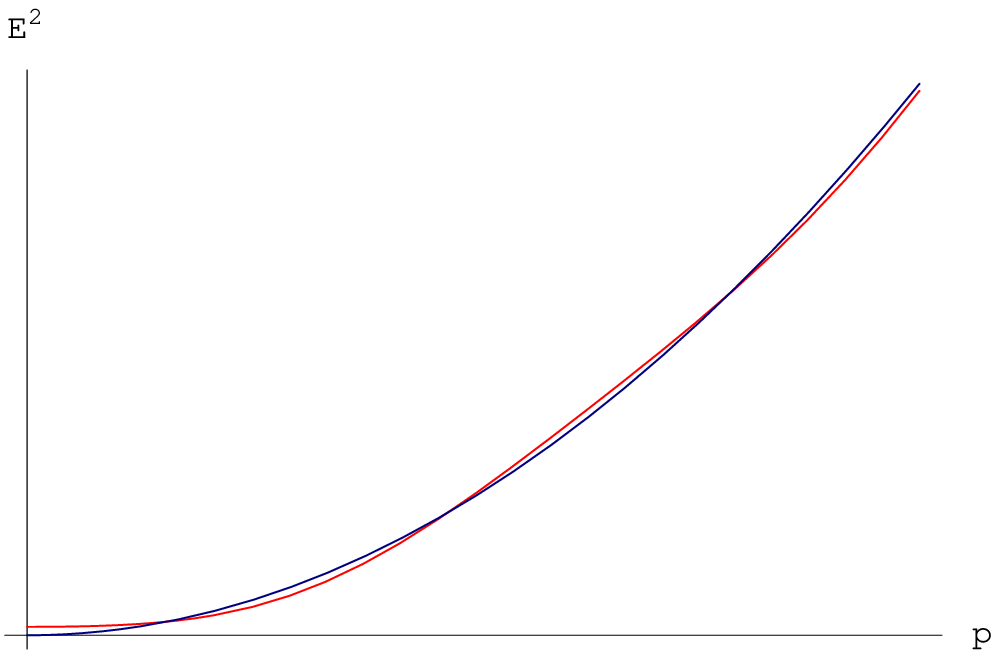}{Dispersion relation for different dipole lengths
in $D=4$, only small wiggling around $E^2=p^2$ is observed.}
\EPSFIGURE{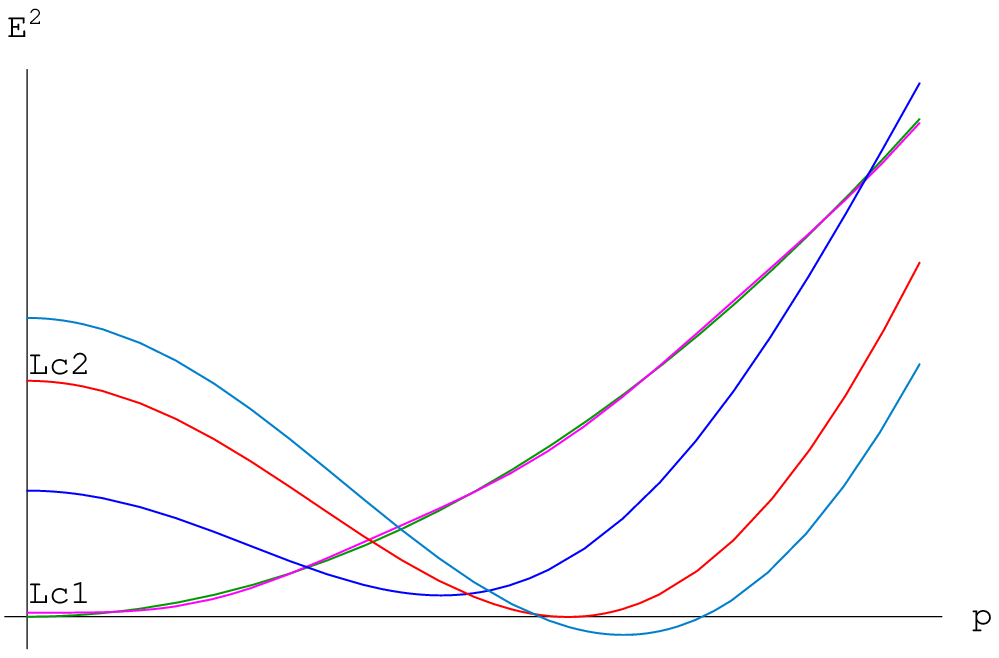}{Dispersion relation for different dipole lengths
in $D=5$. At $L_{c1}$ it develops a minimum away from $p=0$ and at $L_{c2}$ it
touches $E=0$. The effect is more pronounced for smaller $L$.}

Due to the presence of the cosine, the correction term can be negative for some
ranges of momentum. It is natural then to ask if the dispersion relation can
develop a minimum away from the origin in momentum space. We can view this as a
condition on $L$ and define a first critical dipole length as
\begin{eqnarray}
\frac{\partial E^2}{\partial p} = 0 &\rightarrow& \left[ \frac{4 \pi^{D/2}}{g
\Gamma(D/2-1)} L_{c1}^{D-4}\right] \cdot (pL) = \sin(pL) \,,\\
&\Longrightarrow& L_{c1}^{D-4} = \frac{g \Gamma(D/2-1)}{4 \pi^{D/2}} \,.
\end{eqnarray}

Note that at weak coupling this can not be fulfilled in $D=4$! In $D=3$ it is
fulfilled for relatively large $L$. In the massive theory this would still be
approximately valid if we assume the mass to be small such that $L m$ is small,
and the right hand side of the dispersion relation is modified by the addition
of the mass term. In $D=5$ this condition is fulfilled for small $L$. The
five-dimensional theory however is bound to inherit the divergences of the
undeformed theory in the planar sector, and therefore will not be
renormalizable. We can also define a second critical length $L_{c2}$ to be the
value where the right hand side of \erf{eq:dispersion} becomes negative. This
would mean that some of the modes have imaginary energies and therefore show a
tachyonic instability. Again, in the four-dimensional theory this can not happen
at weak coupling. In $D=3$ this happens however for $L \geq L_{c2} \approx
\frac{49}{g}$, whereas in $D=5$ it happens for $L\leq L_{c2} \approx
\frac{g}{308}$. This would imply that the mode $p_{\rm min}$ that minimizes
\erf{eq:dispersion} develops a vacuum expectation value. Since it is a non-zero
momentum mode that condenses the new ground state spontaneously breaks
translation invariance in the direction of the dipole moment! This behaviour is
reminiscent of the behaviour of Moyal deformed $\phi^4$ theory where
it was argued in \cite{Gubser:2000cd} that the UV/IR mixing gives rise to stripe
phases upon adding sufficiently negative tree level mass terms. This was later
on confirmed by lattice studies in \cite{Bietenholz:2004xs}. In our case of the
dipole deformed theory, we expect the results to be qualitatively valid in the
presence of a small positive or negative mass squared term at tree level. It
would be rather interesting to see if these results can be confirmed by a
lattice study of the theory in $D=3$.

\paragraph*{Four-point function.}
Let us have now a quick look to the quantum corrections to the four-point
function. We specialize to $D=4$ and study the possible divergences. The
one-loop correction to the four-point function can be computed from the possible
Wick contractions of
\begin{eqnarray}
\Gamma_{\rm 1-loop}^{(4)}(x_1,x_2,x_3,x_4) &=& \int\dd^4z\,\dd^4w ~\bigg\langle
\phi^\dagger(x_1) \phi^\dagger(x_2) \phi(x_3) \phi(x_4) \times \\
&& \hspace*{2em} \times~ [\lambda (\phi^\dagger \phi)^2(w) + g (\phi^\dagger
\phi)(w+\frac{L}{2})
\,(\phi^\dagger\phi)(w-\frac{L}{2}) ] \times \nonumber\\
&& \hspace*{2em} \left. \times~ [\lambda (\phi^\dagger \phi)^2(z) + g
(\phi^\dagger\phi)(z+\frac{L}{2}) \,(\phi^\dagger \phi)(z-\frac{L}{2}) ]
\right\rangle \nonumber ~.
\end{eqnarray}
The divergences in spacetime arise from the pointwise multiplication of two
propagators
 \begin{equation}
\langle \phi(x) \phi^\dagger(y) \rangle = \Delta(x-y)\; \mathrm{and}\;
\Delta(z)\Delta(z) \approx \log(\Lambda) \delta(z)\;.
\end{equation}
The non-planar contributions amount to multiplication of two propagators with
arguments shifted by multiples of $L/2$, e.g. $\Delta(w-z+L/2) \,
\Delta(w-z-L/2)$. Performing all possible Wick contractions and retaining only
the divergent contributions we find
\begin{eqnarray}
\Gamma_{\rm 1-loop,div}^{(4)} &=& \frac{-1}{8\pi^2} \log(\Lambda)  \int \dd^4z\,
\left\langle \phi^\dagger(x_1) \phi^\dagger(x_2) \phi(x_3) \phi(x_4) \times
\right. \nn \\
&& \times \Big[ (20 \lambda^2 + 2 g^2) (\phi^\dagger \phi)^2(z) + (16 \lambda g
+ 4 g^2) (\phi^\dagger \phi)(z+\frac{L}{2}) (\phi^\dagger\phi)(z-\frac{L}{2}) +
\nn \\
&& \hspace*{2em} \left. + 2 g^2 (\phi^\dagger \phi)(z+L) (\phi^\dagger\phi)(z-L)
\Big] \right\rangle ~.
\end{eqnarray}
This shows that the theory with the $\lambda$ and $g$ interactions in $D=4$ is
strictly speaking not renormalizable. The additional divergence is however
proportional to a dipole star-product term with twice the dipole length. The
divergence can therefore be absorbed into a new tree level term in the action
with star-product and dipole length $2L$. Since this term arises at one loop it
is also natural to assume that its coupling constant is proportional to $g^2$;
its non-planar one-loop contribution to the two-point function can therefore be
neglected to order $g$! It is clear, that at order $g^3$ a new vertex with
star-product structure and dipole length $3L$ will be induced. In general, at
order $g^n$ one needs to assume a tree level Lagrangian of the form
\begin{equation}
{\cal L} = \sum_{k=0}^n g_k \,(\phi^\dagger\phi) \Big(x-k \frac{L}{2} \Big)
\,(\phi^\dagger\phi) \Big(x+k \frac{L}{2}\Big) ~,
\end{equation}
where the couplings can be assumed to obey $g_k \approx g_0^k$ for $k>0$, i.e.
$g_1 = \cO(g_0)$, $g_2 = \cO(g_0^2)$, etc.
Although the theory is not renormalizable in the usual sense, the divergences
stemming from the just discussed corrections to the four-point function are
under control, and can be absorbed in counterterms of star-product form. This
observation has been made already \cite{Dasgupta:2001zu}. 

\subsection{Corrections to KK-masses}  \label{ssec:KKstates}
Let us now consider the dipole scalar field theory  on $\bR^{1,D-2}\times\bS^1$.
The periodicity on the $\bS^1$ direction, with radius $R$, implies that momenta
along that direction are discrete. Since we deal with a complex scalar field
with a $\U(1)$ global symmetry, the field can pick up a phase once transported
around the circle. Thus, if we impose twisted boundary conditions
\be
\phi(x+2\pi R) =\e^{2\pi i\alpha} \phi(x) ~, \quad \alpha\in[0,1) ~,
\ee
the momentum along the $\bS^1$ is
\be
p_{\bS^1} =\frac{n+\alpha}{R} ~.
\ee

From now on, when studying the mass corrections to the KK-states of the tower
coming from the $\bS^1$, we will put the bare mass to zero, $m=0$. We will also
consider the dipole length to be oriented only along the $\bS^1$.

After an appropriate Wick rotation (we assume time to be one of the non-compact
dimensions) the non-planar amplitude takes now the form
\be
\cA_{\rm n-p} =2gT \int\frac{\dd^d k_E}{(2\pi)^d} \sum_{n\in\bbZ} \,\cos\Big(
2\pi TL(n_p-n) \Big) \,\frac{1}{k_E^2 +[2\pi T(n+\alpha)]^2} ~.
\ee
As already stated in the introduction, we used the language of finite field
theory and have set $T=\frac{1}{2\pi R}$. The external momentum is given by
$2\pi T n_p$. The amplitude is of course independent of the inflowing momentum
along the non-compact directions. Note that the argument of the cosine is
independent of the twist parameter $\alpha$.

We notice that $TL \equiv b$ is reduced to the lattice $\left[
-\frac{1}{2},\frac{1}{2} \right)$, because of the periodicity of the cosine
function. Therefore, the amplitude does not care about the magnitude of $TL$
(which can be adjusted by changing $L$), but only about its non-integer part.
Now we split the cosine into exponentials and further perform the
$k$-integration by using the Schwinger parametrization. This yields
\bea
\cA_{\rm n-p} &=& g \,\frac{T^{d-1}}{4\pi} \,\e^{2\pi ibn_p} \sum_{n\in\bbZ}
\int_0^\infty\dd t ~t^{-d/2} ~\e^{-\pi t(n+\alpha)^2 -2\pi ibn} ~+{\rm c.c.} =
\nn \\
&=& g \,\frac{T^{d-1}}{4\pi}
\frac{\Gamma\left(1-\frac{d}{2}\right)}{\pi^{(2-d)/2}} \left\{ \e^{2\pi ibn_p}
\sum_{n\in\bbZ} \frac{\e^{-2\pi inb}}{|n+\alpha|^{2-d}} +\e^{-2\pi ibn_p}
\sum_{n\in\bbZ} \frac{\e^{2\pi inb}}{|n+\alpha|^{2-d}} \right\} ~.\qquad
\eea

We can now evaluate the amplitude using zeta-function techniques. As explained
in the appendix the sums in the amplitude
give a representation of the one-dimensional Epstein zeta function for $2-d >1$.
The amplitude can be regularized 
by analytic continuation of the Epstein zeta function. 
We also make use of the functional identity \erf{eq:functionalid} of the
appendix and can write the result as
\be
\cA_{\rm n-p} =  g \,\frac{T^{d-1}}{2\pi}
\frac{\Gamma\left(\frac{d-1}{2}\right)}{\pi^{(d-1)/2}} \Re\left[ \e^{2\pi
ib(n_p+\alpha)} ~Z\left| \ba{c} b \\ \alpha  \ea \right| (d-1)
\right] ~. \label{eq:bosonic-amp}
\ee

Let us analyse some of the properties of this amplitude. We begin with its
singularity structure. 
From \erf{eq:zetasingstruct} we see that there are two possible pole-like
singularities
\be
\cA_{\rm n-p} = g \,\frac{T^{d-1}}{2\pi}  \cos\Big( 2\pi b(n_p+\alpha)\Big)   
\left( \frac{2\delta_{\alpha,0}}{d-2} -\frac{2\delta_{b,0}}{d-1} +\mbox{two
regular terms} \right) ~.
\ee
The physical meaning of these two poles is the following:
\begin{itemize}
\item for $d=2$ and $\alpha=0$ the pole corresponds to the infrared divergence
of the dimensionally reduced theory in
two dimensions,
\item for $d=1$ and $b=0$  the zeta function only regularizes the infrared
divergence of the theory in $(D=2)$, but leaves us with the ultraviolet one.
\end{itemize}
The presence of the poles can therefore be traced back to the simultaneous
presence of UV and IR divergences.

When the twist parameter $\alpha$ vanishes, the amplitude is factorized as
\bea
\cA_{\rm n-p} &=& g \,\frac{T^{d-1}}{2\pi} \cos(2\pi bn_p) \left\{ \frac{2}{d-2}
-\frac{2\delta_{b,0}}{d-1} \right. \,+ \\
&& \hspace*{2em} +\int_1^\infty\dd t \left[ t^{-d/2} ~\Big( \vartheta_3(b|it) -1
\Big) +t^{(d-3)/2} ~\e^{-\pi b^2t} ~\Big( \vartheta_3(ibt|it) -\delta_{b,0}
\Big) \right] \bigg\} ~. \nn
\eea
If the parameter $b$ vanishes it can be written in terms of generalized Riemann
zeta functions as
\bea
\cA_{\rm p} = g \frac{T^{d-1}}{2\pi}
\,\frac{\Gamma\left(1-\frac{d}{2}\right)}{\pi^{(1-d/2)}} \,[ \zeta_R
(2-d,\alpha) + \zeta_R(2-d,1-\alpha) ] ~,
\eea
which is of course the zeta-function regularized result for the planar
amplitude. For example in $d=3$ and without twist
we get 
\begin{equation}
\cA_{\rm p} = \frac{(2\lambda +g) T^2}{6} ~,
\end{equation}
where we have also taken into account the contributions from the undeformed
vertex with coupling $\lambda$. Sticking for a moment to the interpretation of
$T$ as temperature, this is nothing but the thermal mass of the scalar fields.

\EPSFIGURE{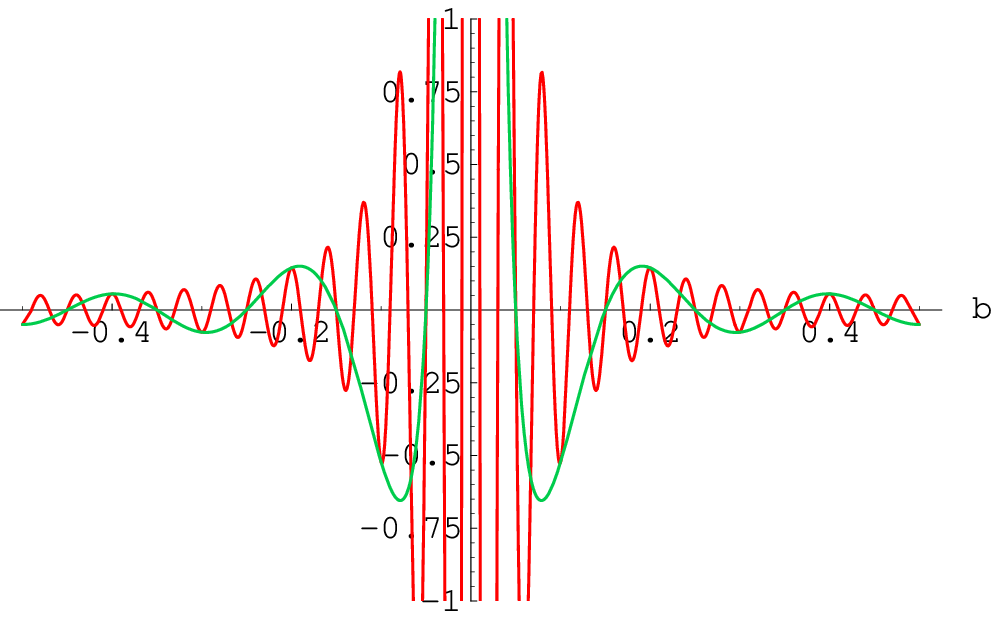}{Non-planar amplitude as a function of the parameter
$b$ in $d=3$ and without twist. The amplitude is plotted for mode number $n=3$
and mode number $n=10$. Although the amplitude is finite at $b=0$ it grows
without bounds for $b\rightarrow 0$!}
The figure shows the behaviour of the non-planar amplitude as a function of $b$
in
$d=3$. The external momentum corresponds to the mode numbers $n_p=3$ and
$n_p=10$. The higher the
mode number is, the faster is the oscillation in $b$. For $b\rightarrow 0$ the
amplitude grows without bounds. However, at precisely $b=0$ the zeta-function
regularization sets in and there the amplitude is finite! The masses of the
KK-states are given by the planar and non-planar contributions
\begin{eqnarray}
M^2_{\rm KK} &=& 4 \pi^2 n_p^2 T^2 + \cA_{\rm p} + \cA_{\rm n-p} = \nn \\[-1ex]
\\[-1ex]
&=& 4\pi^2 n_p^2 T^2 + (2\lambda + g)\frac{T^2}{6} + g \frac{T^2}{2 \pi^2 b^2}
~, \nn
\end{eqnarray}
where in the last line we made an expansion for small $b$, keeping only the
leading term in $d=3$ and without twist. For small $b$ the correction to the
KK-mass is independent from the mode number. Since $b$ can be arbitrarily small,
the non-planar part can give rise to very large KK-masses. Note, however, that
in the case of a commutator interaction at tree level, $g = -\lambda$, the
contribution of the non-planar amplitude is negative! The KK-mass is then
reduced by the non-planar contribution. In fact, for sufficiently small $b$ the
value of $M^2_{\rm KK}$ can become negative for the lowest mode numbers! This
presumably means that the corresponding field in the $d$-dimensional theory
becomes tachyonic and develops a vacuum expectation value!

For $D=4$ ($d=3$), we have seen in our discussion of the non-compact case that
dipole interaction terms with multiples of the dipole length $L$ are necessary.
These interactions give rise to non-planar correction to the KK-masses of the
same form but with $b$ replaced by the reduction $b_k$ to $(-1/2,1/2]$ of $k
TL$. It can occur then that, although $b$ is not small, $b_k$ is small and could
lead to large contributions. We argued that it is consistent to assume that
these interactions are of order $g^k$ but this might not be enough to suppress
the \mbox{non-planar} amplitude, as happens for example by taking $TL=b=0.333$,
$b_3=-0.001$ and $g=0.1$ which gives $g^3/b_3^2 \sim 10^3$!

\section{The `Wess--Zumino' dipole theory}  \label{sec:wzmodel}
In this section we will do the same analysis as in the preceding subsection, but
now for a theory with bosons and fermions. We shall start thus by considering
the Wess--Zumino action in $D=4$.

We need a $\U(1)$ symmetry to define the dipole deformation and we will chose
the $\U(1)_\cR$ of the
supersymmetry algebra. Therefore, we must consider the massless Wess--Zumino
theory. We shall drop also one star-product inside the integral, so that the
propagators of this theory are the same as those in the commutative one. The
action is
\bea
S_{\rm \star WZ} &=& \int\dd^D x \left\{ \partial_\mu\phi^\dag\partial^\mu\phi
-i\psi\sigma^\mu\partial_\mu\overline{\psi}
-g^2\,\phi^\dag\star\phi^\dag\star\phi\star\phi \right.- \nn\\
&& \hspace*{4em} \left. -g\,\phi \star\psi^\alpha \star\psi_\alpha -g\,\phi^\dag
\star\overline{\psi}_{\dot\alpha} \star\overline{\psi}^{\dot\alpha} \right\} ~.
\eea
It is interesting to see explicitly how supersymmetry is broken by the dipole
deformation.
To do so, we go back to the action with the auxiliary fields
\be
S_{\rm \star WZ} =\int\dd^D x \left( \partial_\mu\phi^\dag\partial^\mu\phi
-i\psi\sigma^\mu\partial_\mu\overline{\psi} +F^\dag F +g\Big( \phi\star\phi\star
F -\,\phi\star\psi^\alpha\star\psi_\alpha \Big) +\mbox{h.c.} \right) ~,
\ee
where we see that the correct dipole vectors are $L_\phi=-2L_{\psi}$ and $L_F
=-2L_\phi$. 
Note that upon integrating out the auxiliary fields $F$, $F^\dagger$ we only
generate the quartic
scalar field coupling $\phi^\dagger\star\phi^\dagger\star\phi\star\phi$ !
Plugging in it the following susy variations
\be
\delta_\xi \,\phi = \sqrt{2} \,\xi\psi ~,\quad \delta_\xi \,\psi_\alpha =
i\sqrt{2} \,(\sigma^\mu \,\overline{\xi})_\alpha \,\partial_\mu\phi +\sqrt{2}
\,\xi_\alpha F ~,\quad \delta_\xi \,F = i\sqrt{2} \,(\overline{\xi}
\,\overline{\sigma}^\mu)^\alpha \,\partial_\mu\psi_\alpha ~,
\ee
one obtains
\bea
\delta_\xi \,S_{\rm \star WZ} &=& \int\dd^D x \Big\{
\delta_\xi\,\mbox{(kinetic)} +g\,\delta_\xi\,\phi \star\phi \star F + g\,\phi
\star\delta_\xi\,\phi \star F +g\,\phi \star\phi \star\,\delta_\xi\,F \,- \nn \\
&& \hspace*{4em} -g\,\delta_\xi\,\phi \star\psi^\alpha \star\psi_\alpha -g\,\phi
\star\delta_\xi\,\psi^\alpha \star\psi_\alpha -g\,\phi \star\psi^\alpha
\star\delta_\xi\,\psi_\alpha \Big\} ~. \label{eq:susyvariat}
\eea
It is clear that none of these terms satisfy $\sum L_i=0$, because the
$\cR$-symmetry does not commute with supersymmetry. As a consequence, one cannot
delete one star-product nor use the cyclicity condition in \erf{eq:susyvariat}.
This simply means that the $\U(1)$ used for the dipole deformation cannot be
regarded as the $\U(1)_\cR$-symmetry of a supersymmetric theory: our action is
not the non-commutative Wess--Zumino action, but we will keep this name as it
still describes a theory of bosons and fermions.

The Feynman rules for this theory are
\vspace*{-2em}
\bea
\begin{picture}(50,50)(0,-3)
\ArrowLine(0,0)(50,0)
\Text(-10,0)[c]{$\psi_\alpha$} \Text(60,0)[c]{$\overline{\psi}_{\dot\alpha}$}
\Text(25,10)[c]{$k$}
\end{picture} \qquad &=& \frac{i\,\sigma^\mu_{\alpha\dot\alpha}
\,\bsk_\mu}{\bsk^2-m^2+i\varepsilon} ~, \\
\begin{picture}(50,50)(0,23)
\DashArrowLine(0,50)(25,25){3} \DashArrowLine(0,0)(25,25){3}
\DashArrowLine(50,50)(25,25){3} \DashArrowLine(50,0)(25,25){3}
\Text(4,37)[c]{$k_1$} \Text(3,14)[c]{$k_2$} \Text(47,13)[c]{$k_3$}
\Text(47,37)[c]{$k_4$}
\Text(-8,50)[c]{$\phi^\dag$} \Text(-8,0)[c]{$\phi^\dag$} \Text(57,0)[c]{$\phi$}
\Text(57,50)[c]{$\phi$}
\end{picture} \qquad &=& -ig^2 \,\exp\left( -\frac{i}{2}L_\phi
\cdot(\bsk_1-\bsk_2-\bsk_3+\bsk_4) \right) ~, \\[1em]
\begin{picture}(50,50)(0,23)
\ArrowLine(0,25)(30,25) \DashArrowLine(50,50)(30,25){3}
\DashArrowLine(50,0)(30,25){3}
\Text(15,35)[c]{$k_1$} \Text(48,36)[c]{$k_2$} \Text(48,14)[c]{$k_3$}
\Text(-8,25)[c]{$\phi$} \Text(59,50)[c]{$\psi^\alpha$}
\Text(59,0)[c]{$\psi_\alpha$}
\end{picture} \qquad &=& -ig \,\exp\left( \frac{i}{2}L_\psi \cdot(\bsk_2-\bsk_3)
\right) ~, \\[1em]
\begin{picture}(50,50)(0,23)
\ArrowLine(0,25)(30,25) \DashArrowLine(50,50)(30,25){3}
\DashArrowLine(50,0)(30,25){3}
\Text(15,35)[c]{$k_1$} \Text(48,36)[c]{$k_2$} \Text(48,14)[c]{$k_3$}
\Text(-8,25)[c]{$\phi^\dag$} \Text(59,50)[c]{$\overline{\psi}_{\dot\alpha}$}
\Text(59,0)[c]{$\overline{\psi}^{\dot\alpha}$}
\end{picture} \qquad &=& -ig \,\exp\left( -\frac{i}{2}L_\psi
\cdot(\bsk_2-\bsk_3) \right) \quad \hbox{since} ~L_{\overline{\psi}}=-L_\psi ~.
\eea
\vspace*{2ex} \\

Our aim is to see how the breaking of supersymmetry manifests itself in the
\mbox{non-planar} sector of the theory. As it is well-known, supersymmetry
guaranties the cancellation of quadratic divergences and we will therefore study
these and the corresponding non-planar amplitudes that are regulated by the
dipole length. We will limit ourselves therefore to investigate the one-loop
corrections to the scalar two-point function.

\subsection{1-loop non-planar correction to bosonic propagator} 
\label{ssec:bosonicWZ}

We consider as before the theory on $\cM=\bR^{1,D-2}\times\bS^1$, and also allow
for twisted boundary conditions on the fields
\bea
\phi(x +2\pi R) &=& \e^{2\pi i\alpha} \,\phi(x) ~, \\
\psi(x +2\pi R) &=& \e^{-\pi i\alpha} \,\psi(x) ~,
\eea
where $\alpha = 0$ are supersymmetric boundary conditions whereas $\alpha=1$
corresponds to
the boundary conditions of thermal field theory. We will take the dipole vectors
to be along the $\bS^1$ again.

The bosonic part can be written directly copying from \erf{eq:bosonic-amp},
noticing that $b=-2b_\psi$, and changing appropriately the coupling $(g\to
g^2)$.
On the other hand, taking into account the combinatorics with a minus sign from
the fermion loop and a $\frac{1}{2}$ from the symmetry factor, the non-planar
1-loop fermionic amplitude is
\be
-i\cA_{\rm n-p}^{\rm F} =-2g^2 \int\frac{\dd^D\bsk}{(2\pi)^D} ~\cos\left(
\frac{L_\psi}{R} \,(n_p+2n) \right) \frac{(\bsp +\bsk)\cdot \bsk}{(\bsp +\bsk)^2
\,\bsk^2} ~,
\ee
where again $n_p$ and $n$ are the external and loop momenta along the $\bS^1$.
Also notice that the twist parameter $\alpha$ drops out from the argument of the
cosine. Besides, we have introduced $b_\psi \equiv TL_\psi$ reduced to its
non-integer part, $b_\psi\in\left[-\frac{1}{2},\frac{1}{2}\right)$, because of
the periodicity of the cosine function.

We will make an expansion in $\bsp$, and compute the leading and subleading
contributions,
\be \label{eq:expansion}
\frac{(\bsp +\bsk)\cdot \bsk}{(\bsp +\bsk)^2 \,\bsk^2} =\frac{1}{\bsk^2}
-\frac{\bsp\cdot\bsk}{\bsk^4} +\left( 2\frac{(\bsp\cdot\bsk)^2}{\bsk^6}
-\frac{\bsp^2}{\bsk^4} \right) +\cO(\bsp^3) ~,
\ee
which includes all the superficially divergent terms (since $D=4$ is implied):
quadratic + linear + logarithmic. 
We will evaluate the corresponding non-planar contributions.
After a Wick rotation of the theory, we give next the expression for each term.

\textit{Quadratic part:}
The part of the non-planar amplitude corresponding to a quadratic divergence is
\be
\cA_{\rm n-p}^{\rm F,(2)} =-2g^2T \int\frac{\dd^d k_E}{(2\pi)^d} \sum_{n\in\bbZ}
~\cos\Big( 2\pi b_\psi \,(n_p+2n) \Big) ~\frac{1}{k_E^2
+4\pi^2T^2(n-\alpha/2)^2} ~,
\ee
and is the same as \erf{eq:bosonic-amp} with the substitutions $g\to g^2$,
$\alpha \rightarrow \alpha/2$ and $n_p \rightarrow n_p/2$. So the part of the
combined amplitude corresponding to the quadratic divergences in $D=4$ is
\begin{eqnarray}
\cA_{\rm n-p}^{(2)} = g^2 \frac{T^2}{2\pi^2} \Re \left[ \e^{2 i\pi b_\phi (n_p+
\alpha)} Z\left|\ba{c} b_\phi \\ \alpha  \ea \right| (d-1) - \e^{-2 i\pi b_\psi
(n_p+ \alpha)} Z\left|\ba{c} b_\phi \\ \alpha/2  \ea \right| (d-1) \right]
~.\quad
\end{eqnarray}
The planar contributions are obtained by setting $b_\phi=b_\psi=0$, and cancel
in the case of supersymmetric boundary conditions $\alpha=0$. The non-planar
bosonic and fermionic contributions \textit{do not cancel} each other even in
the case of supersymmetric boundary conditions. This is a direct consequence of
the fact that the amplitudes depend explicitly on the different dipole lengths
related to the $\cR$-charges
through $b_\phi$ and $b_\psi$ respectively.

\textit{Linear part:} We will assume that the $d$-dimensional inflowing momentum
$p$ vanishes, since we want to compute the amputated correction to the two-point
function. Therefore $\bsp =(n_p+\alpha)/R$. The part of the non-planar amplitude
giving the linear divergence thus is
\be
\cA_{\rm n-p}^{\rm F,(1)} =2g^2T \int\frac{\dd^d k_E}{(2\pi)^d} \sum_{n\in\bbZ}
~\cos\Big( 2\pi b_\psi \,(n_p+2n) \Big)
~\frac{(n_p+\alpha)(n-\alpha/2)/R^2}{\Big[ k_E^2 +4\pi^2T^2(n-\alpha/2)^2
\Big]^2} ~.
\ee

In $D=4$, the final expression for this divergence is
\be \label{eq:lineardiv}
\cA_{\rm n-p}^{\rm F,(1)} = -2g^2 T^2 \,(n_p+\alpha) ~\Re\left[ \e^{-\pi ib_\phi
n_p} \left( \frac{1}{2\pi i}\frac{\partial}{\partial b_\phi} +\frac{\alpha}{2}
\right) Z\left| \ba{c} \alpha/2 \\ b_\phi \ea \right| (4-d) \right] ~,
\ee
where all the computations are contained in appendix \ref{app:computationWZ}.
The derivative of the Epstein zeta function must be understood for $b_\phi\neq
0$: one cannot get the result for $b_\phi=0$ as the limit $b_\phi\to 0$ in this
expression. In such a case, one simply has to trace back to the series and see
that it is an alternating sum. Therefore, in that case the linear divergence
vanishes.

\textit{Logarithmic part:}
The last two terms in the power expansion \erf{eq:expansion} give the
logarithmically divergent contributions. As before, we assume that the external
momentum only runs along the circle. After Wick rotation, these terms look like
\bea \label{eq:logdivEuclid}
\cA_{\rm n-p}^{\rm F,(0)} &=& -2g^2T \int\frac{\dd^d k_E}{(2\pi)^d}
\sum_{n\in\bbZ} ~\cos\Big( 2\pi b_\psi \,(n_p+2n) \Big) \times \\
&& \hspace*{4em} \times \left\{ 2\,\frac{(n_p+\alpha)^2(n-\alpha/2)^2/R^4}{\Big[
k_E^2 +4\pi^2T^2(n-\alpha/2)^2 \Big]^3} - \frac{(n_p+\alpha)^2/R^2}{\Big[ k_E^2
+4\pi^2T^2(n-\alpha/2)^2 \Big]^2} \right\} ~, \nn
\eea
which after all computations in appendix \ref{app:computationWZ}, and
specializing to $D=4$, is
\bea \label{eq:logdiv}
\cA_{\rm n-p}^{\rm F,(0)} &=& \frac{g^2T^2}{4} \,(n_p+\alpha)^2 ~\Re\left[
\e^{-\pi ib_\phi n_p} ~Z\left| \ba{c} \alpha/2 \\ b_\phi \ea \right| (4-d)
\right] ~.
\eea

\section{Gauge dipole field theory}  \label{sec:gaugeft}
Finally we want to have a brief look to the dipole deformation of QED. Until now
all the $\U(1)$ symmetries
that we used for the dipole deformations were global symmetries. Now we want to
use the local $\U(1)$ symmetry. Since the gauge field is neutral it has dipole
length zero. The field strength tensor is therefore undeformed. The dipole
deformation allows for three different actions of the gauge group:
\begin{itemize}
\item left action matter fields\\
$ \psi \rightarrow U \star \psi$: \hfill
$D_\mu \psi = \partial_\mu + i g A_\mu \star \psi  =
\partial_\mu + i g A_\mu (x-\frac{L}{2}) \psi (x) $
\item right action matter fields\\
$ \psi \rightarrow \psi \star U^\dagger$: \hfill
$D_\mu \psi = \partial_\mu - i g \psi  \star A_\mu = \partial_\mu - i g
A_\mu(x+\frac{L}{2})\psi (x) $
\item adjoint action matter fields\\
$ \psi \rightarrow U\star\psi \star U^\dagger$: \hfill
$ D_\mu \psi = \partial_\mu + i g [\psi  , A_\mu]_\star =\partial_\mu + i g
\Big( A_\mu(x-\frac{L}{2})\psi-\psi (x) A_\mu(x+\frac{L}{2}) \Big)$
\end{itemize}
Let us chose a commutator-like interaction of the gauge field with the Dirac
spinor $\Psi$. The action is
\begin{equation}
S_{\rm\star QED} = \int \dd^4x\,\left( -\frac{1}{4} F_{\mu\nu}F^{\mu\nu} +
\bar\Psi i \gamma^\mu (\partial_\mu + i g A_\mu \star \Psi -
i g \Psi \star A_\mu ) - m \bar \Psi \Psi\right) \,.
\end{equation} 

The Feyman rule for the interaction vertex is
\vspace*{-1em}\hfill
\[
\begin{picture}(50,50)(0,27)
\Photon(0,30)(60,30){3}{4}\ArrowLine(90,60)(60,30) \ArrowLine(90,0)(60,30)
\LongArrow(20,37)(40,37)
\Text(72,53)[c]{$p$} \Text(72,5)[c]{$q$} \Text(30,45)[c]{$k,\mu$}
\end{picture} \hspace*{4em} \sim \;\;2 i g \gamma^\mu \sin\left(\frac{k\cdot
L}{2}\right) ~.
\]
\vspace*{1em}\hfill \\
The polarization tensor for the gauge field at one loop is
\begin{equation}
\Pi^{\mu\nu} = 4 g^2 \sin\left( \frac{p\cdot L}{2}\right)^2 \int
\frac{\dd^4k}{(2\pi)^4} ~\tr\left[ \gamma^\mu \frac{1}{\slash \hspace{-6pt}k -
m}\gamma^\nu \frac{1}{\slash \hspace{-6pt}k -  \slash\hspace{-6pt} p - m}  
\right] ~.
\end{equation}

We see that the Feynman integral is unchanged compared to the usual QED.
However, the amplitude does
crucially depend on the inflowing momentum through the $\sin^2$ term! This means
that the logarithmic divergence of the integral is multiplied with
$\sin^2(pL/2)$. It is important to realize that this happens here for a
two-point
function. We argued however that the star-product necessarily must have zero
effect on the tree-level two-point functions. Therefore, it is impossible to
absorb the appearing logarithmic divergence in the fields or parameters of the
tree level Lagrangian! The theory is not renormalizable, and contrary to what we
found in the scalar interaction case, there is no star-product term that could
be introduced to deal with this divergence! The problem can be traced back to
the neutrality of the gauge field under the $\U(1)$ symmetry and the commutator
interaction. If instead we use left (or right) multiplication the phases of the
star-product just cancel in the one-loop polarization tensor.

\section{Conclusions}  \label{sec:conclude}
We have reinvestigated certain aspects of dipole deformed field theories. Our
emphasis was on the computation of quantum corrections to the KK-state masses in
a simple compactification. Along the way we also found that dipole scalar field
theory might allow for spontaneous breaking of translation symmetry. Our
analysis was based on a simple one-loop computation and it is not clear if the
properties of the dispersion relation allowing for this phase transition persist
to higher loops or non-perturbative corrections. However, this is an addressable
problem. Lattice studies in the case of the Moyal bracket deformed theory have
shown the formation of stripe phases. It should not be too difficult to
formulate dipole deformed theories on the lattice and perform an analogous
study.

The corrections to the masses of KK-states showed a very interesting pattern.
The dipole length $L$ together with the radius of compactification $R = 1/(2\pi
T)$, forms a dimensionless parameter which we called $b$. It is remarkable that
this parameter is compact, i.e. takes values only in the interval $(-1/2, 1/2]$.
The interesting corrections stem from \mbox{non-planar} graphs, in which the
UV-divergences are regulated by the presence $b$. For $b\rightarrow 0$ the
regularization becomes less effective, and therefore the non-planar contribution
becomes very large and can even overwhelm the tree-level contribution. Depending
on the form of the tree level interaction it might be the case that the
non-planar graph contributes with a minus sign to the square of the KK-mass and
for small enough $b$ the corresponding mode might even become tachyonic!

We also have seen that in the dipole deformation of a supersymmetric theory in
which the $\U(1)_\cR$
symmetry is used for the deformation the supersymmetry is broken. In the planar
sector the quadratic divergences still cancel but in the non-planar sector the
contributions corresponding to quadratic divergences at $b=0$ do not cancel due
to the different dipole lengths of the fields circulating the loop.

Finally, we showed that dipole-deformed QED with adjoint action of the gauge
group is not renormalizable
in a way that would only allow star-product terms in the tree level Lagrangian.
This problem might
be cured only in highly supersymmetric extension like the one based on the
$\cN=4$ theory.

\acknowledgments
The research of K.\,L. is supported by the Ministerio de Ciencia y
Tecnolog\'{\i}a through a Ram\'on y Cajal contract and by the Plan Nacional de
Altas Energ\'{\i}as FPA-2003-02-877. The research of S.\,M. is supported by an
FPI 01/0728/2004 grant from Comunidad de Madrid and by the Plan Nacional de
Altas Energ\'{\i}as FPA-2003-02-877. S.\,M. also wants to thank G. S\'anchez for
her support. K.\,L. would like to thank S. Theisen and C. N\'u\~nez for useful
discussions and the physics departments of Trinity College Dublin, University of
Wales Swansea and Queen Mary University of London where parts of this work have
been presented prior to publication for hospitality.

\appendix

\section{Epstein zeta function}  \label{app:zetafunct}
All of the 1-loop calculations that we perform in the compact case can be given
a closed expression in terms of Epstein zeta functions, a zeta function for
quadratic forms. For a positive integer $p$, let $\vec{g},\vec{h}\in\bbR^p$,
$\vec{n}\in\bbZ^p$. Let us further define the scalar product of any two vectors
in $\bbR^p$ as $(\vec{g},\vec{h})=\sum_{i=1}^p g_i \,h_i$, and the positive
definite quadratic form as
\be
\varphi(\vec x) =\sum_{i,j=1}^p c_{ij} x_i x_j ~,
\ee
where $(c_{ij})$ is called the \textit{module}. The Epstein zeta function of
order $p$ and \textit{characteristic} $\left| \ba{c} \vec{g}\\ \vec{h} \ea
\right|$ is defined as a function of the complex variable $s$ as \cite{epstein} 
\be
Z\left| \ba{c} \vec{g}\\ \vec{h} \ea \right| (s)_\varphi
:=\sum_{\vec{n}\in\bbZ^p}{}^\prime [\varphi( \vec{n} +\vec{g} )]^{-s/2}
~\e^{2\pi i(\vec{n},\vec{h})} ~.
\ee
Notice the prime in the summation, indicating that in case $\vec{g}$ belongs to
the integer lattice one has to subtract the value of $\vec{n}$ such that
$\vec{n} +\vec{g}$ vanishes.

We will focus on $p=1$ with module the identity. Thus $[\varphi(n+g)]^{-s/2}
=|n+g|^{-s}$; hence
\be \label{eq:zetafunc1}
Z\left| \ba{c} g\\ h \ea \right| (s) =\sum_{n\in\bbZ}{}' \frac{\e^{2\pi
inh}}{|n+g|^s} ~.
\ee
Our aim is to give the analytic continuation of this zeta function over the
complex plane. To extract the singular structure one first obtains the integral
representation. Using the Euler Gamma function, $\Gamma(\alpha) =\int_0^\infty
\dd t \,t^{\alpha-1} \,\e^{-t}$; with a change of variables
$\alpha=\frac{s}{2}$, $t=\pi z^2 \,\xi$ and $z=|n+g|$, one arrives at
\be \label{eq:zetafunc2}
Z\left| \ba{c} g\\ h \ea \right| (s) =\frac{\pi^{s/2}}{\Gamma(s/2)}
\sum_{n\in\bbZ}{}' \int_0^\infty\dd\xi ~\xi^{(s-2)/2} ~\e^{-\pi\xi (n+g)^2 +2\pi
inh} ~.
\ee
Now one splits the integration interval as $\int_0^\infty =\int_0^1
+\int_1^\infty$. In the first integral perform a change of variables
$\xi=t^{-1}$ and a Poisson resummation
\be
\sum_{n=-\infty}^\infty \e^{-\pi n^2\tau +2n\pi z\tau} =\frac{\e^{\pi\tau
z^2}}{\sqrt{\tau}} \sum_{m=-\infty}^\infty \e^{-\pi m^2/\tau -2\pi imz} ~.
\ee
Recalling the definition of the Jacobi $\vartheta_3$-function
\be
\vartheta_3 (z|\tau) :=\sum_{n\in\bbZ} \e^{i\pi n^2\tau +2\pi inz} ~,
\ee
we have the following expression for the Epstein zeta function
\bea
Z\left| \ba{c} g\\ h \ea \right| (s) &=& \frac{\pi^{s/2}}{\Gamma(s/2)}
\int_1^\infty\dd t\left\{ t^{(s-2)/2} ~\e^{-\pi g^2t} ~\vartheta_3(h+igt|it) +
\right. \nn \\
&& \hspace*{7em} \left. +t^{-(s+1)/2} ~\e^{-\pi h^2t-2\pi igh}
~\vartheta_3(g-iht|it) \right\} ~.
\eea

Singularities appear in each term for $n=0$ and either $g=0$ or $h=0$. Adding
and subtracting those terms from the summations, and integrating formally yields
the integral representation of the Epstein zeta function
\bea
Z\left| \ba{c} g\\ h \ea \right| (s) &=& \frac{\pi^{s/2}}{\Gamma(s/2)} \left\{
\frac{2\delta_{h,0}}{s-1} -\frac{2\delta_{g,0}}{s} ~+ \right. \nn \\ 
&& \hspace*{4em} +\int_1^\infty\dd t \left[ t^{(s-2)/2} ~\e^{-\pi g^2t} ~\Big(
\vartheta_3(h+igt|it) -\delta_{g,0} \Big) + \right. \label{eq:zetasingstruct} \\
&& \hspace*{8.5em} \left. \left. +t^{-(s+1)/2} ~\e^{-\pi h^2t-2\pi igh} ~\Big(
\vartheta_3(g-iht|it) -\delta_{h,0} \Big) \right] \right\} ~. \nn
\eea
Now it is clear that the function is meromorphic in the whole complex $s$-plane
with a simple pole at $s=1$.

\paragraph{Functional identity.}
Besides the straightforward relation
\be
Z\left| \ba{c} -g\\ -h \ea \right| (s) =Z\left| \ba{c} g\\ h \ea \right| (s)  ~,
\ee
one can obtain the functional identity for the Epstein zeta function, which
interchanges the values of $g$ and $h$. Starting with the integral
representation \erf{eq:zetafunc2}, and after a Poisson resummation and a polar
change of variables $t\to t^{-1}$,
\be \label{eq:functionalid}
\frac{\Gamma\left( \frac{s}{2} \right)}{\pi^{s/2}} ~Z\left| \ba{c} g\\ h \ea
\right| (s) =\e^{-2\pi igh} ~\frac{\Gamma\left( \frac{1-s}{2}
\right)}{\pi^{(1-s)/2}} ~Z\left| \ba{c} h\\ -g \ea \right| (1-s)  ~.
\ee

\section{Explicit computation for dipole WZ}  \label{app:computationWZ}
This appendix shows the calculations performed to get the expressions for the
linear divergence \erf{eq:lineardiv}, and the logarithmic divergence
\erf{eq:logdiv}.

As said in the text, taking the inflowing momentum only along the circumference
gives the following linearly divergent term
\be
\cA_{\rm n-p}^{\rm F,(1)} =2g^2T \int\frac{\dd^d k_E}{(2\pi)^d} \sum_{n\in\bbZ}
~\cos\Big( 2\pi b_\psi \,(n_p+2n) \Big)
~\frac{(n_p+\alpha)(n-\alpha/2)/R^2}{\Big[ k_E^2 +4\pi^2T^2(n-\alpha/2)^2
\Big]^2} ~.
\ee
One can put the cosine function into exponentials, and use the Schwinger
parametrization in the denominator to perform a Gaussian integration. Then
\bea
\cA_{\rm n-p}^{\rm F,(1)} &=& 4\pi^2 g^2 T^3 \,(n_p+\alpha) \sum_{n\in\bbZ}
\Bigg\{ \left(n-\frac{\alpha}{2}\right) \left[ \e^{2\pi ib_\psi n_p} \e^{4\pi
ib_\psi n} +\hbox{c.c.} \right] \times \nn \\
&& \hspace*{1em} \times \int_0^\infty\dd\xi ~\xi \int\frac{\dd^d k_E}{(2\pi)^d}
\e^{-\xi k_E^2 -4\pi^2 T^2\xi(n-\alpha/2)^2} \Bigg\} ~.
\eea
Performing the Gaussian integration, and further doing $t=4\pi T^2\xi$ yields
\bea
\cA_{\rm n-p}^{\rm F,(1)} &=& g^2 T^{d-1} \,(n_p+\alpha) \sum_{n\in\bbZ} \Bigg\{
\left(n-\frac{\alpha}{2}\right) \left[ \e^{2\pi ib_\psi n_p} \e^{4\pi ib_\psi n}
+\hbox{c.c.} \right] \times \nn \\
&& \hspace*{2em} \times \int_0^\infty\dd t ~t^{(2-d)/2} \e^{-\pi
t(n-\alpha/2)^2} \Bigg\} ~.
\eea
Using the Euler Gamma function as we did in appendix \ref{app:zetafunct}, and
recalling $b_\phi=-2b_\psi$, one arrives at
\be
\cA_{\rm n-p}^{\rm F,(1)} = 2g^2 T^{d-1} \,(n_p+\alpha) ~\frac{\Gamma\left(
2-\frac{d}{2} \right)}{\pi^{(4-d)/2}} \Re\left[ \e^{-\pi ib_\phi n_p}
\sum_{n\in\bbZ} \left(n-\frac{\alpha}{2}\right) \frac{\e^{-2\pi ib_\phi
n}}{|n-\alpha/2|^{4-d}} \right] ~.
\ee
This expression can be given formally in terms of the Epstein zeta function and
its derivative with respect to $b$
\be
\cA_{\rm n-p}^{\rm F,(1)} = -2g^2 T^{d-1} \,(n_p+\alpha) ~\frac{\Gamma\left(
2-\frac{d}{2} \right)}{\pi^{(4-d)/2}} \Re\left[ \e^{-\pi ib_\phi n_p} \left(
\frac{1}{2\pi i}\frac{\partial}{\partial b_\phi} +\frac{\alpha}{2} \right)
Z\left| \ba{c} \alpha/2 \\ b_\phi \ea \right| (4-d) \right] ~.
\ee

Now we do the computation for the logarithmically divergent terms. As before, we
can put the cosine as exponentials in \erf{eq:logdivEuclid}, and use the
Schwinger trick to get, after Gaussian integration
\bea
\cA_{\rm n-p}^{\rm F,(0)} &=& -\frac{4\pi^2 g^2T^3}{(2\sqrt\pi)^d}
\,(n_p+\alpha)^2 \Bigg\{ \left( \e^{-\pi ib_\phi (n_p+2n)} +\hbox{c.c.} \right)
\times \nn \\
&& \hspace*{2em} \times \int_0^\infty\dd\xi ~\xi^{-d/2}
\Big(4\pi^2T^2\xi^2(n-\alpha/2)^2 -\xi\Big) ~\e^{-4\pi^2 T^2\xi(n-\alpha/2)^2}
\Bigg\} ~.
\eea
Performing the change of variables $t=4\pi T^2\xi$, and mapping to the Euler
Gamma function as with the linear divergence, one obtains
\be
\cA_{\rm n-p}^{\rm F,(0)} =-\frac{g^2T^{d-1}}{4} \,(n_p+\alpha)^2 \left(
1-\frac{d}{2} \right) \frac{\Gamma\left( 2-\frac{d}{2} \right)}{\pi^{(4-d)/2}}
\sum_{n\in\bbZ} \left( \e^{-\pi ib_\phi (n_p+2n)} +\hbox{c.c.} \right)
\frac{1}{|n-\alpha/2|^{4-d}} ~,
\ee
and after some algebra it can be given in terms of the Epstein zeta function
\be
\cA_{\rm n-p}^{\rm F,(0)} =-\frac{g^2T^{d-1}}{4} \,(n_p+\alpha)^2 \left(
1-\frac{d}{2} \right) \frac{\Gamma\left( 2-\frac{d}{2} \right)}{\pi^{(4-d)/2}}
\Re\left[ \e^{-\pi ib_\phi n_p} Z\left| \ba{c} \alpha/2 \\ b_\phi \ea \right|
(4-d) \right] ~.
\ee


\end{document}